\documentclass[a4paper,11pt]{elsarticle}

\usepackage{amsmath}     
\usepackage{graphicx}
\usepackage{amsfonts}
\usepackage{amssymb}
\usepackage{amsthm}
\usepackage[pdftex,colorlinks,linkcolor={blue},citecolor={red}]{hyperref}
\usepackage{color}
\usepackage{float}
\usepackage{verbatim}
\usepackage{indentfirst}

\catcode `\@=11
\@addtoreset{equation}{section}

\catcode`\@=12

\def\psihat{\hat{\psi}}
\def\bvec{\vec{b}}
\def\evec{\vec{e}}
\def\kvec{\vec{k}}

\def\xvec{\vec{x}}

\def\etavec{\vec{\eta}}

\def\alphatilde{\tilde{\alpha}}

\def\tautilde{\tilde{\tau}}
\def\uvectilde{\tilde{\vec{u}}}
\def\Rbb{\mathbb{R}}

\def\Zbb{\mathbb{Z}}
\def\Ccal{\mathcal{C}}
\def\nn{\nonumber}
\newtheorem{Proposition}{Proposition}[section]
\def\nn{\nonumber}
\newcommand{\ud}{\,\mathrm{d}}
\newcommand{\be}{\begin{equation}}
\newcommand{\en}{\end{equation}}

\newcommand{\bea}{\begin{eqnarray}}
\newcommand{\ena}{\end{eqnarray}}

\newcommand{\beano}{\begin{eqnarray*}}
\newcommand{\enano}{\end{eqnarray*}}

\newcommand{\bei}{\begin{itemize}}
\newcommand{\eni}{\end{itemize}}

\newcommand{\ip}[2]{\langle {#1}\, | \,{#2}\rangle}
\newcommand{\CC}{\mathcal C}
\newcommand{\R}{\mathbb R}%\input{alphabet}
\begin{document}
\begin{frontmatter}

\title{A spatio-temporal Gaussian-Conical wavelet with high aperture selectivity for motion and speed analysis}

\author{Patrice Brault$^{\rm a}$\footnote{IEEE Senior Member; {\it E-mail address}: patrice.brault@lss.supelec.fr}
and Jean-Pierre Antoine$^{\rm b}$\footnote{{\it E-mail address}: Jean-Pierre.Antoine@uclouvain.be}}
\address{$^{\rm a}$ \emph{\small LSS, Laboratory of Signals and Systems, CNRS - Supelec - Paris-Sud University,\\
Gif-sur-Yvette, France}
\\[2mm]
$^{\rm b}$ \emph{\small Institut de Recherche en Math\'ematique et Physique, Universit\'e catholique de Louvain \\
B - 1348   Louvain-la-Neuve, Belgium}}
%
% ----------------------------------------------------------
\begin{comment}
\twoauthors
{P. Brault, IEEE Senior Member}
{LSS,\\
CNRS - Supelec - Paris-Sud University\\
Gif sur Yvette, France}
%
%\twoauthors
%
{J-P. Antoine}
{Inst. Rech. Math. Phys. (IRMP)\\
UCL, Louvain-La-Neuve, Belgium}
\end{comment}
% ----------------------------------------------------------

%=============== Begin Abstract =================
%
\begin{abstract}
The construction of a spatio-temporal wavelet and its tuning to speed was first realized in the 90s on the Morlet wavelet by M. Duval-Destin \cite{Duval-Destin91a,Duval-Destin92}.
This enabled to demonstrate the capacities of the speed-tuned Morlet for psychovisual analysis. This construction was also used very efficiently in a powerful aerial target
tracking algorithm by Mujica et al.\cite{Mujica99,Mujica2000}. In the last decade, this tool was proposed as an elegant and
efficient alternative framework to the Optical Flow (OF), the Block Matching (BM) or the phase difference, for the study of motion estimation in image sequences.
Nevertheless, the aperture selectivity of the 2D+T Morlet wavelet presents some difficulties.
Here we propose to replace the 2D Morlet wavelet by a Gaussian-Conical (GC) wavelet for the spatial part of the spatio-temporal wavelet, since
the GC wavelet has a better  aperture selectivity and allows a very simple adjustment of the aperture.
Therefore we build a new, highly directional, speed-tuned wavelet called
Gaussian-Conical-Morlet (GCM) wavelet. Like the speed-tuned 2D+T Morlet, the new wavelet presents very good characteristics in motion estimation and tracking,
namely long temporal dependence, robustness to noise and to occlusions, and supersedes the OF (Optical Flow) and BM (Block Matching) techniques.
However, for aperture selectivity, directional speed-capture and spectral recognition and tracking, GCM easily outperforms Morlet. This paper describes the GCM construction, 
utilization and aperture performances.
\end{abstract}

\keyword
Extraction of motion parameters \sep speed tuning \sep continuous wavelet transform \sep    directional and conical wavelets \sep angular and aperture selectivity
 \endkeyword
\end{frontmatter}
%----------------------------
\section{Introduction}
%----------------------------
The continuous wavelet transform has proved to be a very efficient tool for signal analysis. In the late 80s and the 90s, developments to adapt the wavelet transform to various motions
 have been proposed by M. Duval-Destin and  R. Murenzi  \cite{Duval-Destin92}. % (see also \cite[Chap.10]{Antoine04b}).

The group of analysis parameters, i.e.,  usually  position,   scale and rotation, has been extended to  speed, acceleration and deformation. This has led to various types of time dependent
 wavelets \cite[Chap. 10]{Antoine04b}.
Then a very performant algorithm for missile tracking was set up by Mujica  et al. \cite{Mujica99,Mujica2000} using such wavelets (see also \cite[Chap.10]{Antoine04b}).
In \cite{Hong08,Wang011a}, the authors use the same  energy-density based algorithm, but with  Expectation-Maximization plus Gaussian mixture approach
 and scale functional relation plus ST processing blocks.
Later, in \cite{Wang011b}, the second named authors also use image transformation to time-varying (1D+T) signals.

The advantage of velocity detection with the motion-tuned spatio-temporal CWT over other known methods, like Optical Flow (OF) \cite{Bernard99.0}, Block Matching (BM)
 and phase difference has been already discussed in \cite{Brault03,Brault05}.
These methods work on the motion of pixels or of blocks, but not on regions or objects. They assume that the object is constant from frame to frame and that the \textit{object signature}
does not change with time. They are not inherently scalable either. Like BM, OF has a short time dependence, which is not very accurate for slow motion or trajectory estimation.
The four characteristics of object  tracking, by spectral signature or spatial scale, long temporal dependence, robustness to noise and robustness to occlusions, are the strength of
wavelet analysis, and we plan to show it further for pertinent feature extraction and recognition in sequence analysis, for object tracking, for video compression
 and for video data mining.
% endred

Because of its compactness both in position space and in frequency space, but also because the symmetry of its envelope, the Morlet wavelet was first chosen to be tuned to speed.
In Fourier space, this wavelet reads as
\vspace*{-1mm}
\begin{equation}  \label{eq:morlet}
\widehat\psi_{\scriptscriptstyle M}(\vec k)  = \sqrt{\epsilon}\,
         \Big(\exp(-{\textstyle\frac 12}|A^{-1}(\vec k - \vec k_0)|^2)
   -  \exp(- {\textstyle\frac 12}|A^{-1} \vec k_{0}|^2)\, \exp(-{\textstyle\frac 12}|A^{-1}\vec k|^2) \Big),
\vspace*{-2mm}
\end{equation}
where $A =\mbox{diag}[1, \epsilon^{-1/2}], \epsilon \geqslant 1,$ is a  $2 \times 2$ anisotropy matrix
 and the correction term, which ensures admissibility of the wavelet, is usually dropped (see \cite[Eq(3.18)]{Antoine04b}), since it is negligible for practical values of the parameters.

Nevertheless  numerous difficulties remain when using this wavelet in directional analysis. Although it has a good capability for directional filtering, its aperture selectivity is poor.
It is \textit{directional} in the sense defined in \cite{Antoine99a} and \cite[Sec.3.3]{Antoine04b}, namely,
\textit{``A wavelet $\psi$ is said to be directional if the effective support of its Fourier transform $\widehat\psi$ is contained in a convex cone in spatial frequency space''},
but the anisotropy parameter $\epsilon>1 $ is needed in order to get a decent angular selectivity.
In addition, the Morlet wavelet has a major drawback: its angular selectivity  increases with the length of the wave vector $\vec{k_0}$, since the support cone gets narrower,
 but at the same time the amplitude decreases as $\exp(-|\vec{k_0}|^2)$.

 {In order to achieve a more efficient directional wavelet, a better method is to consider a smooth function with support in a strictly convex cone $\mathcal{C}$ and behaving inside
 this cone as $P(\vec{k})e^{-\vec{\zeta}\cdot  \vec{k}}$ where $\vec{\zeta} \in \mathcal{C}$
and $P(\cdot)$ is a polynomial. This leads to the conical wavelets, in particular the  Cauchy  wavelet and the Gaussian-conical wavelet,
if the exponential is replaced by a Gaussian  \cite{Antoine99a,Antoine04b}. These are genuine directional wavelets that   don't suffer from the defects of the Morlet wavelet. }
We have used these wavelets as a basis for a new construction of motion-tuned, and in particular speed-tuned wavelets.
The development of  these wavelets  and their use in motion analysis  is the aim of the present paper.

\section{Preliminaries: the 2D continuous WT}
%-------------------------------------------------------
In order to motivate our construction and to fix notations, we begin with a brief reminder of the 2D continuous wavelet transform (CWT), following \cite[Chap.2]{Antoine04b}.
A 2D wavelet is a function $\psi \in L^2(\Rbb^2,\ud \vec x)$ satisfying the {admissibility} condition
\be
c_{\psi}\equiv (2\pi)^2 \int_{\mathbb{R}^2}\, \ud\kvec \,\frac{|\psihat(\kvec)|^2}{|\kvec|^2} < \infty,
\en
where $\psihat$ is the Fourier transform of $\psi$. In practice, this condition is often replaced by the slightly weaker one
\be \label{eq:admiss2}
\widehat\psi(\vec{0})=0  \;\Longleftrightarrow\; \int_{\mathbb{R}^2}  \ud\xvec \;\psi(\xvec)=0.
\en
Given a 2D signal (an image) $s \in  L^2(\Rbb^2)$, its CWT with respect to the wavelet $\psi$ is given by the inner product
\begin{align}
W_{\psi}s (\bvec,a,\theta) &=  \ip{\psi_{\bvec,a, \theta}}{s}\nonumber
\\
&=
a^{-1}\int_{\mathbb{R}^2}  \ud\xvec \;\overline{\psi(a^{-1}r^{-\theta}(\xvec-\bvec)}\, s(\xvec), \label{eq:cwt-position}
\\
&=
a \int_{\mathbb{R}^2}  \ud\kvec \;e^{\mathrm{i}\bvec.\kvec} \,\overline{\widehat \psi(a r{-\theta}(\kvec)}\, \widehat s(\kvec), \label{eq:cwt-freq}
\end{align}
where the overbar denotes   complex conjugation, $\psi_{\bvec,a, \theta}$  is a copy of $\psi$ translated by $\bvec\in \mathbb{R}^2$,
 dilated by a factor $a > 0$, and rotated by an angle $\theta\in[0,2\pi]$, that is,
\begin{equation}
\psi_{\bvec,a, \theta}= {a}^{-1}\,\psi\, ( {a}^{-1}\,r_{\theta}^{-1}(\xvec-\bvec)),
\end{equation}
 {where $r^\theta$ is the familiar $2\times 2$ rotation matrix of angle $\theta$
\begin{align*}
r^{\theta}= \left(
\begin{array}{cc}
\cos\theta & -\sin\theta \\
 \sin\theta & \cos\theta
\end{array}
\right)
\end{align*}
}
 Inverting the transform, we obtain the reconstruction formula
\be
s(\vec x) = c_{\psi}^{-1}\, \iiint  \ud \vec b\;\frac{\ud a}{a^{3}} \, \ud\theta \; \psi_{\vec b,a,\theta}(\vec x) \;W_{\psi}s(\vec b,a,\theta).
\label{reconst}
 \en
These formulas will be extended to the 2D+T case in Section \textcolor{red}{\ref{SecMotionTuning}} below.
% endred

Next comes the choice of wavelet. Two possibilities are available:

(i) \emph{Isotropic wavelets}, i.e., rotation invariant wavelets, well adapted for a pointwise analysis;

(ii) \emph{Anisotropic wavelets: }when the aim is to detect oriented features in an image, which is the case here,
  one has to use  a wavelet which is \emph{not} rotation invariant.
 The best angular selectivity will be obtained if  $\psi$ is \emph{directional}  \cite{Antoine96a,Antoine99a}, which means that the (numerical) support of $\widehat\psi$  in spatial
frequency space is contained in a convex cone with apex at the origin (that is, that the wavelet is numerically negligible outside the   cone).
 When  this condition is not respected, as in the anisotropic Mexican hat wavelet, whose `footprint'  is an ellipse centered at the origin, the directionality of the wavelet becomes very poor.
A typical directional wavelet is the 2D Morlet wavelet \eqref{eq:morlet}. % or the conical wavelets.

%-------------------------------------
\section{Conical wavelets}
\label{sec-conical}
%--------------------------------------
However, some directional  wavelets (e.g. Morlet) have a good capability of directional filtering, but their angular selectivity remains poor.
Thus was introduced the concept of a \emph{conical} wavelet \cite{Antoine99b},
which generalizes the 1D Cauchy wavelet of Paul (see \cite[Sec. 2.1]{Antoine99a}), namely,
\be
\widehat\psi_m(\omega) = \left\{ \begin{array}{ll}
0,                            & \mbox{for} \; \omega<0,\\
\omega^m \, e^{-\omega},  & \mbox{for} \; \omega \geq 0.
\end{array}\right.
\label{eq:paul}\en
In 1D, the positive half-line is a convex cone. Thus a natural generalization to 2D (and in fact to $n$D) will be
   a smooth   function $\psihat^{\mathcal{C}}(\vec k)$, with support included in a strictly convex cone
$\mathcal{C}$ with apex at the origin, and which behaves inside this cone as
 $P(k_1,k_2,...,k_n)e^{-\zeta.\kvec}$,  where $\zeta \in \mathcal{C}$ and $P $ is an $n$-variable polynomia

By analogy with \eqref{eq:paul}, the radial behavior in $e^{-\zeta.\kvec}$ of this conical wavelet gives it its specific name  of   \emph{Cauchy wavelet}.
In order to get a better radial localization, the slowly decreasing exponential term can be replaced by a Gaussian in $k_x$ \cite{Vandergheynst98,Antoine99b}.
 We will use the term \emph{Gaussian-Conical} for this specific case, that we will discuss in Section \ref{subsec-GC}.

%-------------------------------------------------
\subsection{The 2D Cauchy wavelet}
%-------------------------------------------------
We take first the 2D  Cauchy wavelet, following \cite[Sec. 3.3.4]{Antoine04b}.
For simplicity, we consider a strictly convex cone, symmetric with respect to the positive $k_x$-axis, namely
$$
\CC := \CC (-\alpha,\alpha) =
\{ \vec{k} \in {\R}^2 \, | \, -\alpha \leqslant \arg \vec{k} \leqslant \alpha,\, \alpha < \pi/2\},
$$
that is, the convex cone determined by the  unit vectors $\vec e_{-\alpha}, \vec e_{\alpha}$.
The dual cone,  with sides perpendicular to those of the first one, is  also convex and reads:
$$
\widetilde{\CC} = \CC (-\tilde{\alpha},\tilde{\alpha}) =
\{ \vec{k} \in {\R}^2 \, | \, \vec{k} \cdot \vec{k'} > 0, \,
\, \forall \,\vec{k'}  \in \CC(-\alpha,\alpha)\},
$$
where $ \tilde\alpha =  -\alpha + \pi/2$.
Therefore $\vec e_{-\alpha}\cdot\vec e_{\tilde\alpha} = \vec e_{\alpha}\cdot\vec e_{-\tilde\alpha} = 0$.
Thus $\alpha$ is the aperture of the cone $\CC$, $\widetilde\alpha$ the  aperture of the dual cone

Given the fixed vector $\vec\eta = (\eta, 0), \eta >0,$ along the axis of the cone,
 we first define the   {Cauchy wavelet} in spatial  frequency variables:
\be \label{cauchy}
{\widehat\psi}^{\;C}_{lm}(\vec k) =
\left\{
\begin{array}{l}
(\vec k \cdot \vec e_{\tilde\alpha})^l \;
(\vec k \cdot \vec e_{-\tilde\alpha})^m \; e^{-\vec k \cdot \vec\eta  },
\quad \vec k \in  {\CC}(-\alpha,\alpha)\;   \\
\, 0, \quad \mbox{otherwise}.
\end{array}
\right.
\en
The Cauchy wavelet $\widehat{\psi}^{\;C}_{lm} (\vec{k})$ is strictly supported in the cone ${\CC} (-\alpha,\alpha)$ and the parameters
$ l,m \in {\mathbb N}^{*}, l, m \geqslant 1$, give the number of vanishing moments of $\widehat\psi$ on the edges of the cone,
and thus control the regularity of the wavelet. One mostly uses the symmetric version of the wavelet, with $l=m$.
The advantage of this wavelet is that its expression in $\xvec$-space can be computed analytically  \cite[Eq.(3.29)]{Antoine04b}, but we won't need it in the sequel.

We note in passing that the Cauchy  wavelets have minimal uncertainty, in the sense that they minimize the product of the variances (uncertainties) of a pair of noncommuting elements of the
underlying Lie algebra of the similitude group. This is the exact equivalent of the usual  minimal uncertainty relation $\Delta p \Delta x \geqslant \hbar/2$ familiar in quantum mechanics,
or  $\Delta B(\omega) \Delta x(t) \geqslant1$, for a signal, with $B(\omega)$ the bandwidth.
We refer to  \cite[Sec. 8.2]{Antoine04b} for a discussion of this topic

%------------------------------------------------------------
\subsection{The 2D Gaussian-Conical wavelet (GC)}
\label{subsec-GC}
\begin{figure}[h]
\begin{center}
\includegraphics[width=6cm,height=4.8cm]{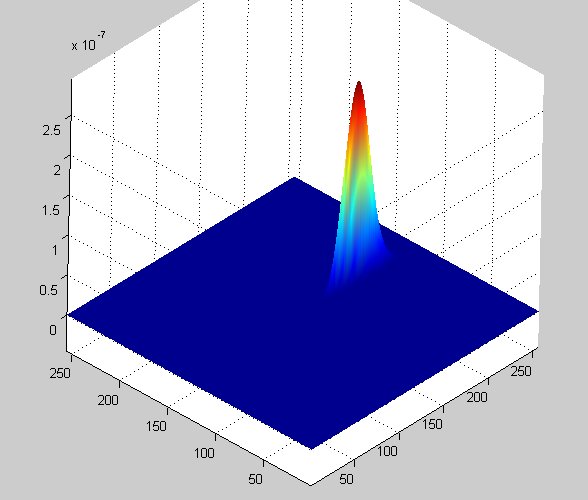}
\hspace{0.2cm}
\includegraphics[width=6cm,height=4.8cm]{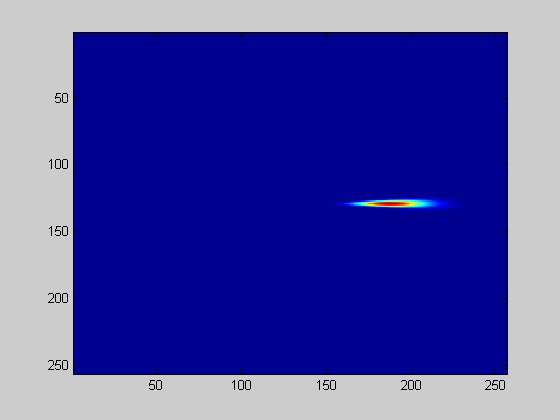}
 \caption[]{\label{ConicalMorlet2D} \small The 2D Gaussian-Conical filter in the $(k_x,k_y)$ plane: the ``shark'' wavelet; side and top views.
The parameters are $\alpha = 10^\circ, l=m=4$.}
\end{center}
\end{figure}

%------------------------------------------------------------
By definition, the 2D Cauchy wavelet has the property that its opening angle $\alpha$ is \textit{totally controllable} \cite{Antoine04b}, independently of the amplitude,
thus it avoids the drawback of Morlet mentioned above.
Nevertheless, although it has a good \textit{angular selectivity}, its \textit{radial selectivity} is poor because
 the exponential term decays slowly as $|\kvec|\rightarrow \infty$. This is why this exponential is often replaced by a Gaussian along $k_x$,
which concentrates the wavelet on its central frequency $(\sqrt{l+m},0) $ \cite{Vandergheynst98,Antoine99b,Jacques04}.
Then $\sigma > 0$ controls the scale localization of the Gaussian. A \textit{center correction term}, $\chi(\sigma)= \sqrt{l+m}\frac{\sigma -1}{\sigma}$,
controls the radial support of $\psi$.
Thus we obtain the expression of the Gaussian-Conical or GC wavelet, in   frequency space \cite[Eq.(3.37)]{Antoine04b}:
%----------------------------
\begin{equation}
{  \widehat\psi^{\,GC}_{lm}(\kvec)= \left\lbrace
\begin{array}{ll}
(\kvec\cdot\evec_{-\widetilde\alpha})^l (\kvec\cdot\evec_{\widetilde\alpha})^m e^{-\frac{\sigma}{2}
 (k_{x} - \chi(\sigma))^2}, \; \kvec \in \Ccal,\\
0,\;\mbox{otherwise}.
\end{array}\right.
}
\end{equation}
The GC wavelet is shown in Fig. \ref{ConicalMorlet2D}.%
%
% SHARK FIGURE : Figure 1
%--------------
%

%
%----------------------------------------------------------------------------
\section{ The new 2D+T, motion-tuned, Gaussian-Conical-Morlet (GCM) wavelet}
%-------------------------------------------------------------------------
\subsection{Construction of the GCM wavelet} 
%-------------------------------------------------------------------------
Building a wavelet for directional velocity analysis, in a sequence of images, starts by building a
spatio-temporal 2D+T wavelet.
Our purpose is to analyze motions in very precise directions and with easy adjustment of this direction. Because the 2D Gaussian conical filter has good capacities
 in spatial resolution and orientation, as well as aperture selectivity, it has been chosen for the spatial component of new velocity-tuned wavelet.

As for the 1D temporal direction,  using a conical wavelet makes no sense for velocity tuning, since a 1D convex cone is a half-line, as in \eqref{eq:paul}.
On the other hand, with the Duval-Destin-Murenzi (DDM) wavelets \cite{Duval-Destin91a,Duval-Destin92}, the Morlet wavelet has demonstrated
 its ability to be shaped and elongated both in the spatial and  the temporal directions,
thus ensuring the adaptation to a psychovisual behaviour as well as the tuning to speed. Because of its Gaussian shape, this wavelet behaves like an ellipsoid.
Elongating this ellipsoid in the vertical or horizontal direction (i.e., along $k_t$ or $k_x$) is easily done by means of the speed-tuning operator introduced in Section \ref{SecMotionTuning}.
Furthermore, the displacement along an hyperboloid of velocities ensures speed capture by ``catching'' the inclination of the Fourier spectrum of a signal for a very specific speed,
thus the speed of this signal. And the behavior of the Morlet ellipsoid with speed tuning perfectly fits the slope of the spectrum: a horizontal (in the $\kvec$ direction)
 ellipsoid for low speeds and a vertical (in the $\omega$ direction) ellipsoid for high speeds.
For this reason we make the choice of a Morlet filter along the temporal wave-vector axis.

As for the DDM case, the resulting 2D+T wavelet is constructed in a separable way in frequency space: the velocity-tuned 2D+T filter is  simply the product of the 2D Gaussian-Conical wavelet
with the 1D Morlet wavelet.We call the resulting wavelet \emph{2D+T Gaussian-Conical-Morlet}  (2D+T GCM, or simply GCM). It is given in Fourier space as
\begin{equation}\label{GCM-develop-0}
{\widehat\psi^{\,GCM}_{lm} (\kvec, \omega)=\left\lbrace
\begin{array}{lll}
\underbrace{\widehat\psi^{\,GC}_{lm}(k_x,k_y)}_{\text{2D Gaussian-Conical}} \cdot \underbrace{\widehat\psi^M(\omega)}_{\text{1D Morlet}},
\; \kvec \in \Ccal (-\alpha,\alpha),\\
\\
0,\; \mbox{otherwise},
\end{array}\right.}
\end{equation}
with $\widehat\psi^M(\omega) = e^{-\frac12(\omega-\omega_{0})^2}$  and $\omega_{0}$ the central frequency of the Morlet wavelet.
Explicitly, the  expression of GCM, in Fourier, is given by
\begin{equation}\label{GCM-develop}
{ \psihat^{\,GCM}_{lm} (\kvec,\omega)=\left\lbrace
\begin{array}{lll}
(\kvec\cdot \evec_{-\alphatilde})^l \,(\kvec\cdot \evec_{\alphatilde})^m \,
e^{-\frac{\sigma}{2} (k_{x} - \chi(\sigma))^2} \, e^{-\frac12(\omega-\omega_{0})^2},  \; \kvec \in \Ccal (-\alpha,\alpha),\\
0,\; \textrm{otherwise.}
\end{array}\right.}
\end{equation}

Note that more sophisticated spatio-temporal wavelets may be designed. They are adapted to a specific relativity group (Galilei, Poincar\'e) 
and are derived from a representation of the 
latter \cite[Sec. 15.3]{aag_book}.  For simplicity, however, and for the sake of comparison with previous works on motion analysis,
we restrict ourselves to the separable case.

Correspondingly, we   consider (2+1)-dimensional signals $s$ (image sequences) of finite energy
$s \in L^{2}(\R^{2}\times \R,{\ud}{\vec x}\, \ud t)$:
\begin{equation}
\|s\|^2 = \iint_{\R^{2}\times \R} \ud\vec x  \,\ud t\;|s(\vec x,t)|^{2} < \infty \,.
\end{equation}
The Fourier transform of $s$ is defined, as usual, by
\begin{equation}
\widehat s (\vec k,\omega) =  {(2\pi)^{-3/2}}\iint _{\R^{2}\times \R} \ud\vec x \: \ud t\;
e^{-i(\vec k\cdot\vec x + \omega  t)}s(\vec x,t).
\end{equation}
where $\vec k $ is the spatial frequency and $\omega $ is the temporal frequency. 

%------------------------------------------------------------------------
\subsection{Tuning the GCM wavelet to motions}
\label{SecMotionTuning}
%------------------------------------------------------------------------

Several motion operators can be applied to the ``mother'' wavelet $\psi$, namely, scaling ($D$), translation ($T$), rotation ($R$) and  speed tuning ($\Lambda$).
The action of these operators on the wavelet is given by
\begin{align}\label{transforms}
[\widehat{D}^{a_s,a_t} \: \widehat{\psi}](\vec{k},\omega)  &= a_s a_t^{1/2} \widehat{\psi}(a_s\vec{k},a_t\omega), \nn\\
[\widehat{T}^{\vec{b},\tau} \: \widehat{\psi}](\vec{k},\omega)  &=  e^{-i(\vec{k}\cdot\vec{b}+\omega\tau)} \widehat{\psi}(\vec{k},\omega), \\
[\widehat{R}^{\theta} \: \widehat{\psi}](\vec{k},\omega) & = \widehat{\psi}(r^{-\theta}\vec{k},\omega),  \nn \\
[\widehat{\Lambda}^c \; \widehat{\psi}](\vec{k},\omega)&= \widehat\psi(c^{q}\vec{k},c^{-p}\omega).\nn
\end{align}
As explained in \cite[Sec.10.3]{Antoine04b}, these operations form a group, with group parameters $g=\{\vec{b},\tau, \theta ; a_s,a_t,c\}$.
Notice that we consider here different scaling parameters in space and time,  $a_{s}$ and $a_{t}$, respectively.
Applying these operations on the wavelet $\psi$, we obtain the dilated, speed tuned wavelet denoted $\psi_{\vec{b},\tau, \theta ; a_s,a_t,c}$.
In terms of the latter, we may now define the 2D+T continuous WT of the signal $s \in L^{2}(\R^{2}\times \R,{\ud}{\vec x}\, \ud t)$ as
\begin{align}
{\mathcal W}_{\psi}s (\vec{b},\tau, \theta ; a_s,a_t,c) &= \ip{\psi_{\vec{b},\tau, \theta ; a_s,a_t,c}}{s}Ê\nn \\
&=  a_s^{-1}a_t^{-1/2}\;\iint \ud\vec k \;\ud\omega \;
e^{-i(\vec k\cdot\vec b + \omega\tau)}\;
\overline{\widehat{\psi}(a_{s}c^{q}\, r^{-\theta}\vec k, a_{t}c^{-p}\omega)}
 \; \widehat s(\vec k,\omega).
\label{eq:2DTCWT}
\end{align}

Like for the speed-tuned Morlet wavelet, the application of these operators to the GCM wavelet is done separately on the spatial 2D conical filter and the temporal 1D Morlet filter,
depending on the temporal or the spatial nature of the transformation.
In particular, the speed tuning is realized by the deformation of the space-time domain with speed. The spatial wave-number (or position) is multiplied by a factor proportional to the speed,
and the temporal wave-number is multiplied by a factor inversely proportional to the speed. These factors are adjusted by the two speed exponents $p= 2/3$ and $q= 1/3$ whose values
are fixed by the constraints on the transformation $\Lambda^c$ \cite{Duval-Destin92}.

After implementation of the transforms defined in \eqref{transforms},
 we can give the expression of the 2D+T GCM in terms of the group  parameters $g=\{\vec{b},\tau, \theta ; a_s,a_t,c\}$:
\begin{equation}\label{GCM-full-parameters}
\begin{array}{ll}
\psihat^{\,GCM}_{\vec{b},\tau, \theta ; a_s,a_t,c} (\kvec,\omega)=
\left\lbrace
\begin{array}{lll}
a_s^{-1}a_t^{-1/2}e^{-i(\vec{k}\cdot\vec{b}+\omega\tau)} \,(c^{q} \,  a_s \, r^{-\theta} \kvec \cdot \evec_{\widetilde\alpha})^l
(c^{q} \, a_s r^{-\theta} \kvec \cdot \evec_{-\widetilde\alpha})^m
 \\[1mm]
\hspace*{1cm}\times\;\exp[-\frac{\sigma}{2} (a_s c^{q} k_{x} - \chi(\sigma))^2]  \exp[-\frac12(c^{-p} a_t  \omega-\omega_{0})^2], \quad \kvec \in \Ccal(-\alpha,\alpha),
\\
0,\; \mbox{otherwise}.
\end{array}\right.
\end{array}
\end{equation}

\begin{figure}[t]
\begin{center}
\includegraphics[width=6cm,height=4.8cm]{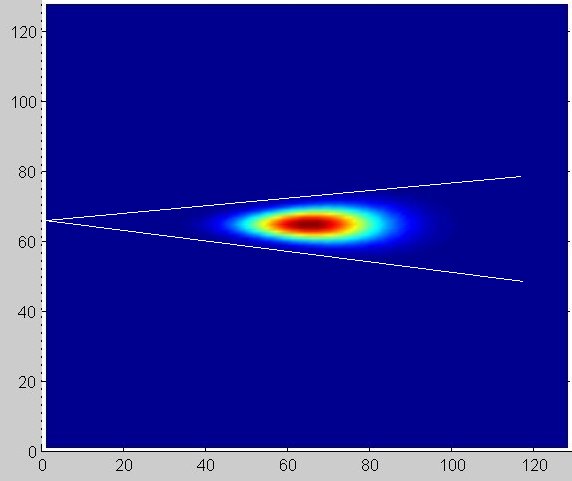}
\hspace{0.2cm}
\includegraphics[width=6cm,height=4.8cm]{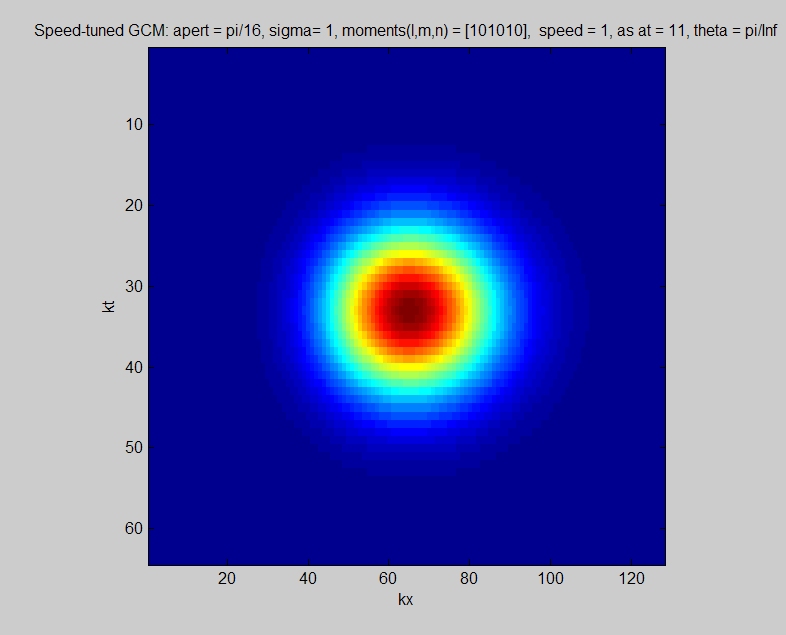}\\
\vspace{0.4cm}
\includegraphics[width=6cm,height=4.8cm]{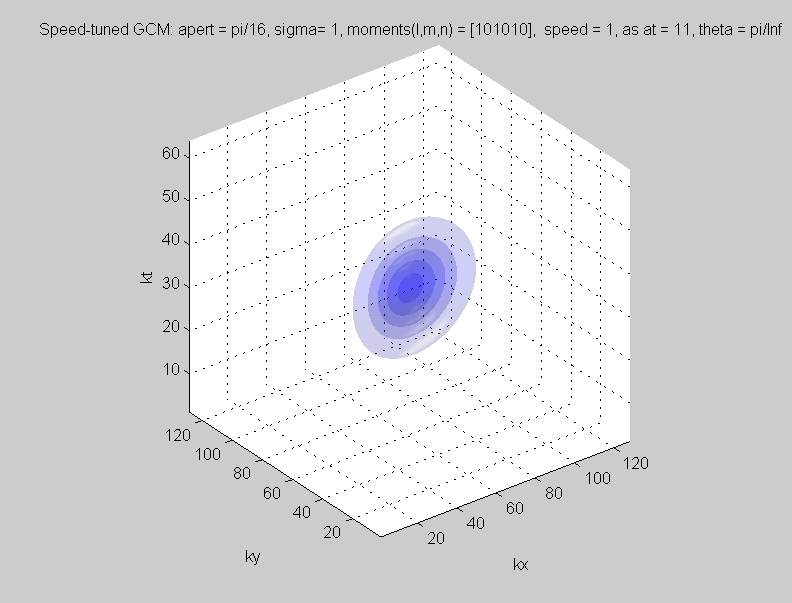}
\hspace{0.2cm}
\includegraphics[width=6cm,height=4.8cm]{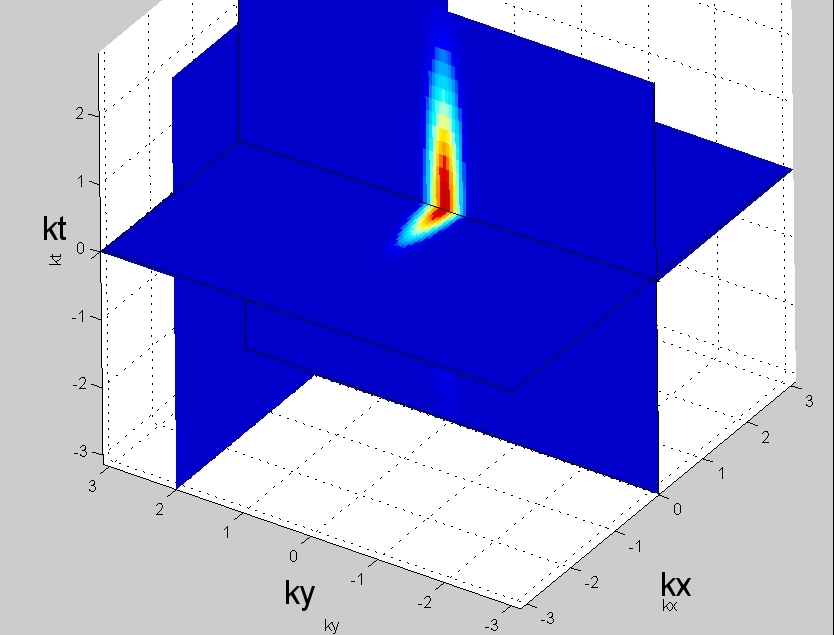}
\caption{\label{ConicalMorlet2DT}\small The 2D+T Gaussian-Conical-Morlet (GCM) wavelet tuned to speed:
(a)  Top view in the plane $(k_x,k_y)$ for $c=1$ (i.e. this is the mother wavelet) and aperture $\alpha=\pi/16$;
(b)  Lateral   view of the same in the plane $(k_x, \omega)$;
(c)  3D view of the GCM  envelope in Fourier space $(k_x,k_y, \omega)$ for $c=1$;
(d) 3D sectional views in $(k_x,k_y,\omega)$ showing the Gaussian amplitude behaviour on the (vertical) Morlet part, as well as on the (horizontal) conical part, and for $c = 2$.}
\end{center}
\end{figure}

\begin{figure}[h]
\begin{center}
\includegraphics[width=4.5cm]{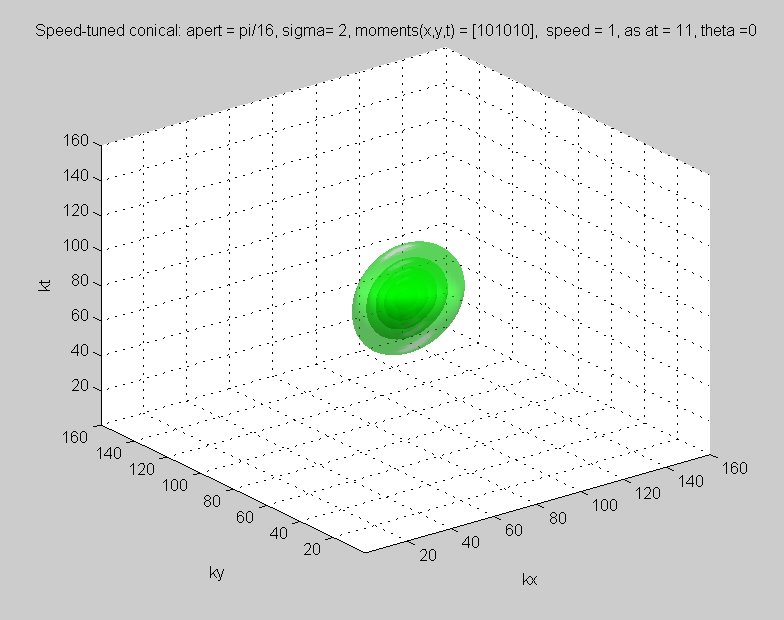}\hspace{0.3cm}
\includegraphics[width=4.5cm]{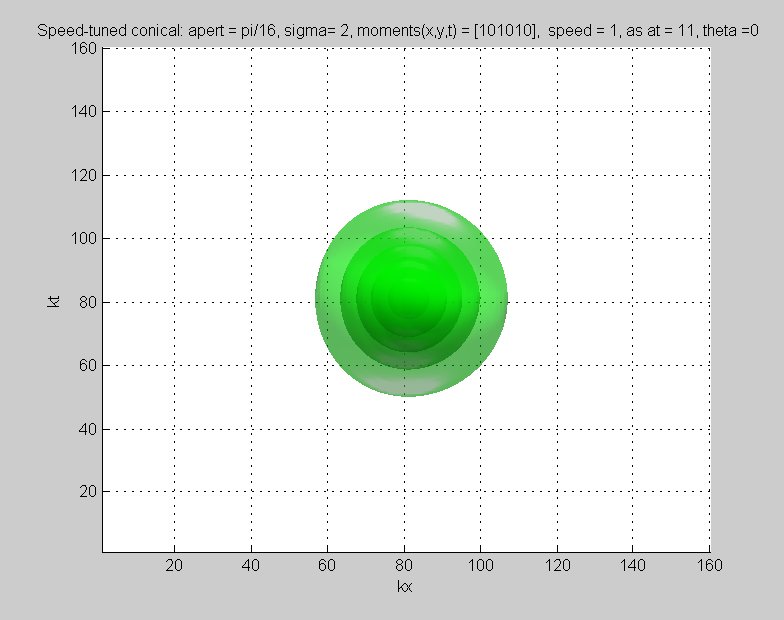}\hspace{0.3cm}
\includegraphics[width=4.5cm]{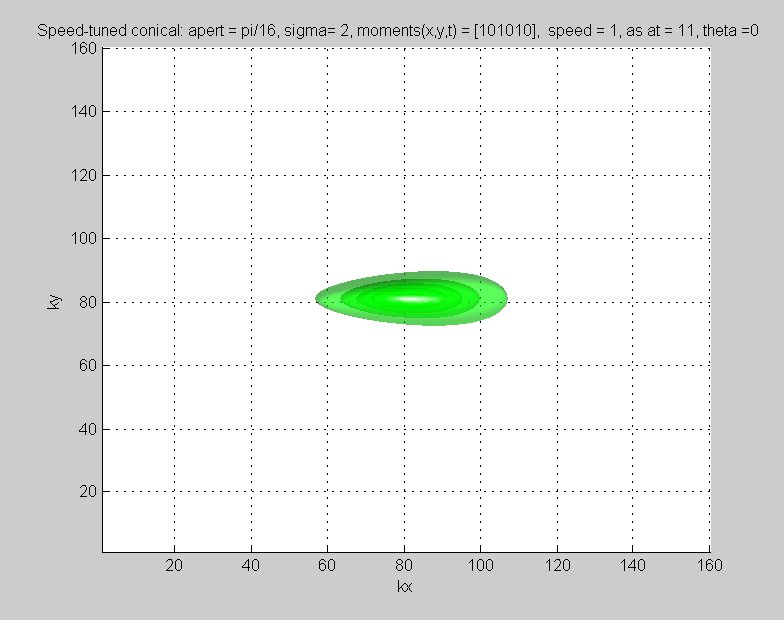}\\
\vspace{0.5cm}
\includegraphics[width=4.5cm]{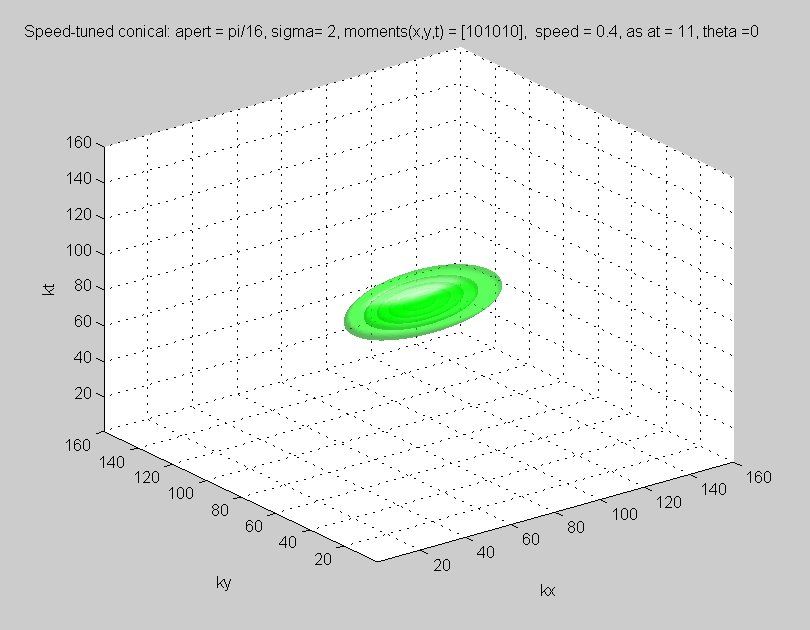}\hspace{0.3cm}
\includegraphics[width=4.5cm]{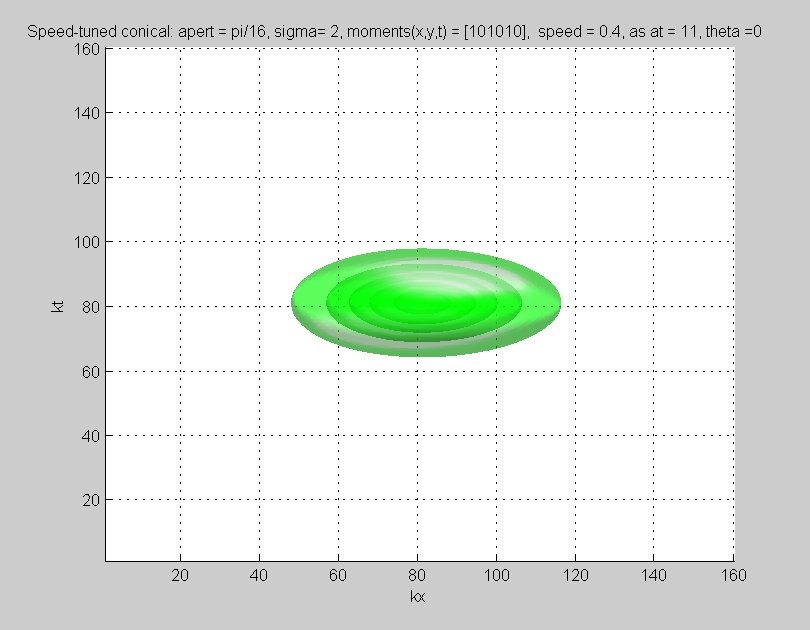}\hspace{0.3cm}
\includegraphics[width=4.5cm]{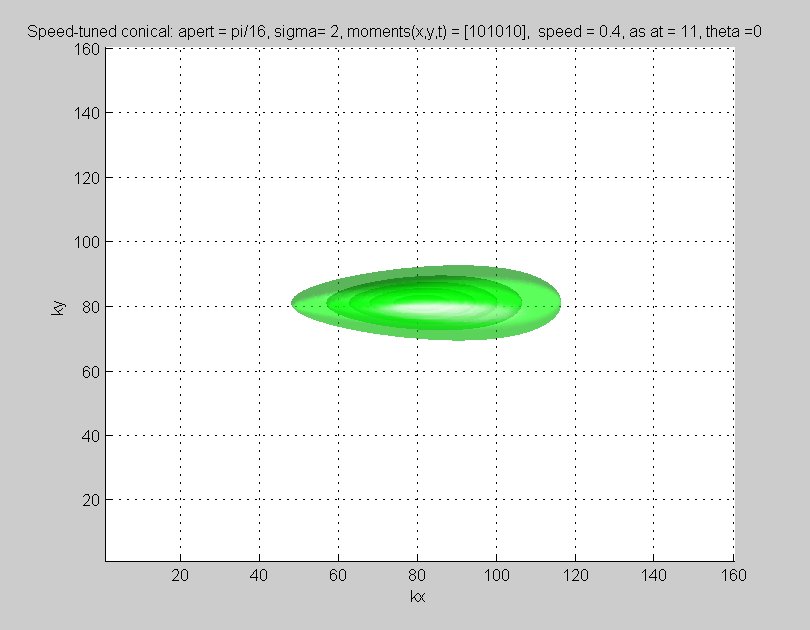}\\
\vspace{0.5cm}
\includegraphics[width=4.5cm]{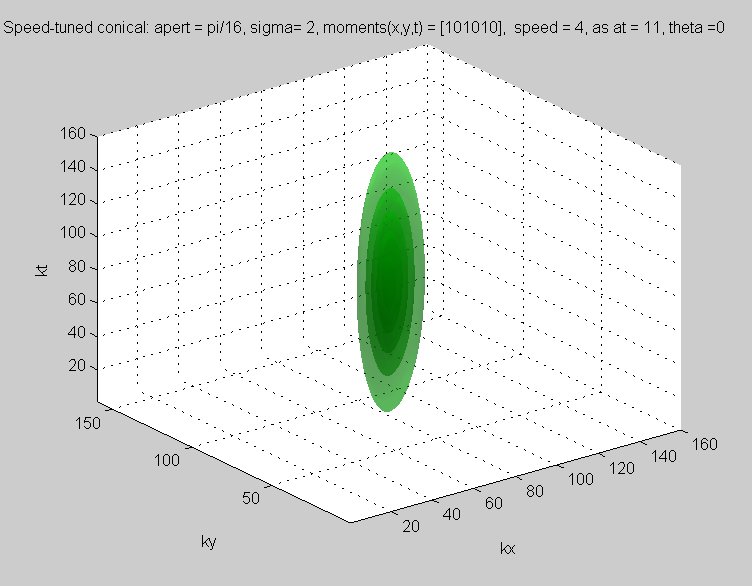}\hspace{0.3cm}
\includegraphics[width=4.5cm]{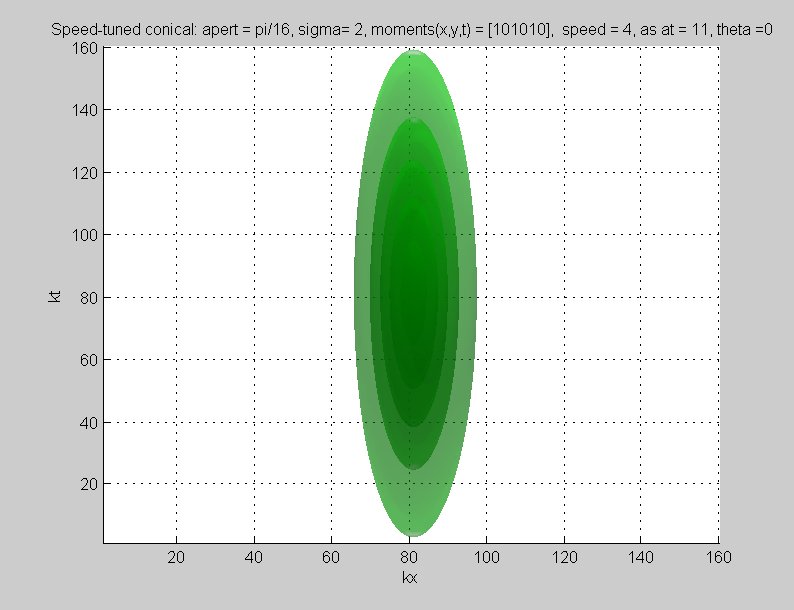}\hspace{0.3cm}
\includegraphics[width=4.5cm]{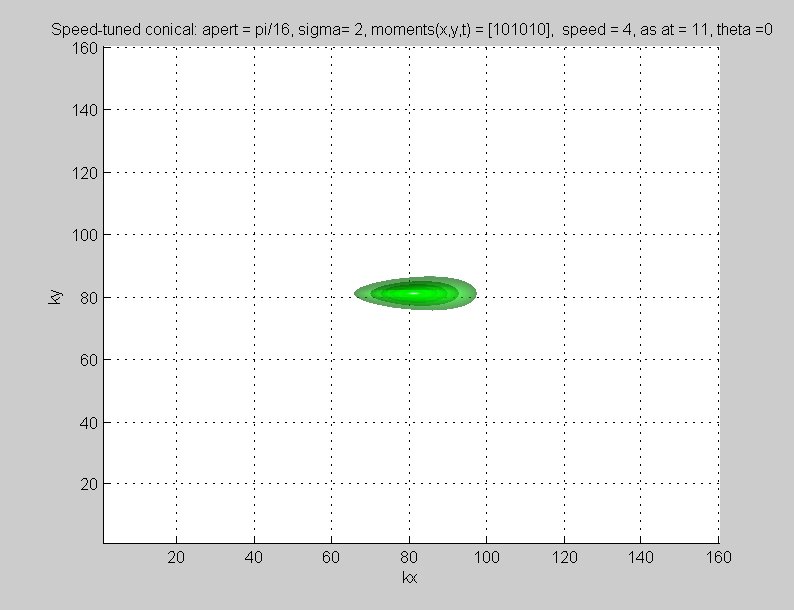}\\

\caption[]{\label{GCM_speed_tuning} \small The  ``progressive'' wavelet GCM  tuned to various velocities with 3D view (left),
side view (center) in $(k_{x},\omega)$ and top view (right) in $(k_{x},k_y)$:
(a)(b)(c) This is the mother wavelet for speed analysis, i.e. with velocity tuning $c=1$ (shown here for sake of comparison with two other velocity tunings below).
The other parameters are: $\sigma=1 ; l=m=10; a_{s}=a_{t}=1; \alpha= \pi/16; \theta=0$; (d)(e)(f) $c=$ 0.4 pix/fr, $\sigma=2$;  (g)(h)(i) $c=$ 4 pix/fr, $\sigma=2$.
 We notice on (f) and (i) how the (spatial) frequency size is reduced with speed. This demonstrates the good psychovisual behavior of wavelets with speed.
We know that the higher the speed the larger the pattern size must be. Here, the wavelet proportionally shrinks in Fourier space (thus widens in position space)
 with an increase of speed, in order to capture higher speeds.}
\end{center}
\end{figure}

{The 2D+T GCM wavelet tuned to speed $c=1$, i.e., the mother wavelet for speed analysis, is shown in Fourier space $(k_x,k_y,\omega)$ 
 in Fig. \ref{ConicalMorlet2DT} (and its sectional view at $c=2$, for sake of clarity). In Fig. \ref{GCM_speed_tuning}, we show the same mother GCM wavelet together with speed-tunings at low ($c=0.4$) and high ($c=4$) speeds. This figure perfectly describes the similarity of psychovisual behavior, with speed variations, between GCM and 2D+T Morlet.}

From here on, our notation will be the following:
The wavelet tuning speed is $c$,   the speed  of the real object in the sequence is  $v_r$ and $v_m$
will be the speed measured, on the energy curve, by the speed-tuned wavelet family.

% Figure 2

%-------------------------------
%Figure 3

%================================================

\subsection{Central frequency and speed capture initialization}
%--------------------------------------------------------------

The GCM wavelet can be transformed to a simple low-pass filter by centering the wavelet in the Fourier plane. This is done by translating the wave-vector $\kvec$ of the conical component
by the value $\kvec_0$ (see below) of its central frequency, and by cancelling the term $\omega_{0}$ for the Morlet part.
With this condition, the conical and the Morlet wavelets do not oscillate any more, in fact they are no longer wavelets, but simply filters.
They also cannot be tuned to scale. The advantage is that all the ``GCM speed-tuned filters'' so obtained are centered in the Fourier plane and can easily capture the initial speed, due to their shape,
 independently of the object scale.
The value of the new central wave-vector of our speed-tuned 2D+T GCM can simply be deduced from the static version. The GCM wavelet has been constructed with axis $\etavec$
in the same direction as $k_x$. Thus the ``static'' central wave-vector corresponds to a maximum of the wavelet along the $k_x$ axis. A derivation of $\psi^{GCM}$ leads in a straightforward manner
to the following value of  the ``static'' central wave-vector: $ k_{x0}=\sqrt{l+m}$  \cite{Jacques04}. If we remark that, in the speed-tuned GCM, the $\kvec$ vector is multiplied by $c^{q}$,
we can thus write the central wave-vector for speed-tuning
\begin{equation}\label{CentralFreqForSpeedTuning}
k^c_{x0}=\frac{\sqrt{l+m}}{c^{q}}
\end{equation}
Because the value of $k_0$ is also affected by all the transformations that the wavelet undergoes, we thus compensate
the conical wave-vector $\kvec$ by the central frequency computed for all transforms, which gives
\begin{equation}
\vec k_0 = \frac{1}{a_{s}} \frac{1}{a_{t}} \frac{\sqrt{l+m}}{c^{q}} (\cos\theta, \sin\theta)
\end{equation}
The expression of GCM centered in   Fourier space becomes
\begin{equation}\label{GCM-full}
\widehat{\Lambda}^g \psihat^{GCM} (\kvec)=
\left\lbrace
\begin{array}{lll}
a_s^{-1}a_t^{-1/2}e^{-i(\vec{k}\cdot\vec{b}+\omega\tau)} \, (c^{q} a_s   r^{-\theta}(\kvec -\kvec_{0})\cdot \evec_{\alphatilde})^l
\; (c^{q}  a_s   r^{-\theta}  (\kvec -\kvec_{0})\cdot \evec_{-\alphatilde})^m
\\
\hspace*{8mm}\times\;\exp[{-\frac{\sigma}{2} ((k_{x}-k_{x0}) - \chi(\sigma))^2}] \exp[{-\frac12(c^{-p}  a_t   (\omega-\omega_{0}))^2}], \quad  \kvec \in \Ccal (-\alpha,\alpha),
\\
0,\; \mbox{otherwise}
\end{array}\right.
\end{equation}
where, as before, the group parameters are $g=\{\vec{b},\tau, \theta ; a_s,a_t,c\}$.
\subsection{Angular resolving power}
%===================================
If the wavelet $\psi$ has its effective support, in spatial frequency, in a vertical cone of aperture $\Delta \phi$, corresponding to $\kvec_0=(0,k_0)$,
the wavelet is considered as concentrated in an ellipse. Its angular resolving power (ARP) is defined by considering the tangents to that ellipse
\cite[Sec. 3.4.1]{Antoine04b}. For Morlet, and for $k_0\gg 1$, this gives
\begin{equation}
ARP(\psi^M)=2 \cot^{-1}(k_0 \sqrt\epsilon)
\end{equation}
which means that the angular resolving power of Morlet only depends on the product $k_0 \sqrt\epsilon$, a result first proposed in \cite{Antoine93}.
For the symmetric Cauchy  or the conical wavelet in general, with support in the cone $\Ccal(-\alpha, \alpha)$, the ARP is simply its opening angle, or aperture, $2 \alpha$.

\subsection{Admissibility, frame bounds and reconstruction}
%==============================================================
Frame theory addresses the issue of reconstructing a signal $s(t)$ from sample coefficients of a particular transform. In general, a family $\psi_{n}$ in $L^2(X)$
is a frame if there exist $A>0$ and $B<\infty$ such that for, all $s \in L^2(X)$~:
\begin{equation}
A||s||^2 \leqslant \sum_{n} |\langle \psi_{n},s \rangle|^2 \leqslant B ||s||^2,
\end{equation}
where $A$ and $B$ are the \textit{frame bounds}.

Most reconstruction algorithms from frame coefficients are of the recursive type or require the calculation of a dual frame, recursive as well.
These algorithms generally require that the frame bounds be known by advance. The convergence rate also depends on the ratio $B/A$ which is an important issue in the use of the
spatio-temporal CWT transform; as is known \cite{Daubechies82, Mallat98}, the closer $B/A$ to $1$, the faster the convergence. We are interested here in knowing the values of the parameters
that yield   fast convergence with the GCM CWT transform.
 {In 1D, estimates for the frame bounds have been given by Daubechies \cite[Proposition 3.3.2]{Daubechies82}.
The same problem has been studied by Murenzi in the 2D case (see \cite[Section 2.4.2]{Antoine04b})  and
by Mujica   for the 2D+T case of DDM wavelets  \cite{Mujica99}. We briefly recall some important steps of
 this issue and the  conditions under which the family considered  is   a frame,   presents a stable reconstruction and   provides fast convergence rates.}

 {The first step is to discretize all wavelet parameters. Let us put
$$
a= a_0^l, \; c = c_0^{n}, \; \theta = q\theta_0, \; a_0, c_0 > 1, \theta_0 = \pi/q_1, \textrm{with}\; q_1\in \Zbb, l,n, q\in \Zbb.
$$
Then, one defines the following quantities:}
\begin{align}
\Lambda(\kvec, \omega) &= \sum_{l,n,q\in \Zbb}|\psihat(a_0^l c_0^{n/3} r^{-\theta_0q}\kvec, \; a_0^l c_0^{-2n/3} \omega )|^2,\\
\Gamma(\bvec, \tau) &= \sup_{\kvec, \omega} \sum_{l,n,q\in \Zbb}|\psihat(a_0^l c_0^{n/3} r^{-q\theta_0}\kvec), \; (a_0^l c_0^{-2n/3} \omega)|^2
& &|\psihat(a_0^l c_0^{n/3} r^{-q\theta_0}(\kvec-\bvec), a_0^l c_0^{-2n/3}(\omega -\tau ))|,\\
\gamma &= \sum_{\substack{m_x,m_y,p\in \Zbb \\ \neq (0,0,0)}} \sqrt{\Gamma(\uvectilde_{m_x, m_y, \tautilde_p}) \Gamma(-\uvectilde_{m_x, m_y, -\tautilde_p})},
\end{align}
with
 {$\uvectilde_{m_x,m_y}=\left[ \frac{2\pi m_x}{b_{x_0}}\frac{2\pi m_y}{b_{y_0}} \right]^T$,
$\tautilde_p = \frac{2\pi p}{\tau_0}$, %$a_0, c_0 > 1, \theta_0 = \pi/q_1$, with $q_1\in \Zbb;$;
and $b_{x_0}, b_{y_0},\tau_0 >0$.}

%%%%%%%%%%%%%
\vspace{.5cm}

With these quantities, the extension of Daubechies's result    to 2D+T frame bounds becomes

%-------------------------------
\begin{Proposition}[Estimates of the frame bounds for the 2D+T case]
If $\psi_d$, $a_0$, $c_0$ and $\theta_0$ are such that~:
\begin{equation}\label{Lamba}
\Lambda_- = \inf_{\kvec, \omega} \Lambda(\kvec, \omega) > 0, \qquad
\Lambda_+ = \sup_{\kvec, \omega} \Lambda(\kvec, \omega) < \infty,
\end{equation}
and if $\Gamma(\bvec, \tau)$ decays fast enough, then there exists $b_{x_0*}$, $b_{y_0*}$, $\tau_{0*} > 0$ such that $\psi_d$ represents a frame for all $0<b_{x_0}< b_{x_0*}$,
$0<b_{y_0}< b_{y_0*}$, $0<\tau_{0}<\tau_{0*}$. Moreover, the following are frame bounds for $\psi_d$~:
\begin{equation}
A = \frac{(2\pi)^{3/2}}{b_{x_0}b_{y_0}\tau_{0}} (\Lambda_{-} - \gamma), \qquad
B = \frac{(2\pi)^{3/2}}{b_{x_0}b_{y_0}\tau_{0}} (\Lambda_{+} + \gamma),
\end{equation}
\end{Proposition}
%--------------------------------

With an appropriate choice of  $a_0$, $c_0$ and $\theta_0$, the values of the infimum and supremum $\Lambda_-$ and $\Lambda_+$, the bounds $A$ and $B$ can be close to each other.
 The parameters $b_{x_0}$, $b_{y_0}$ and $\tau_0$ are also adjusted so that  $\gamma$ is close to 0. Then the ratio $B/A$ can be close to one, thus resulting in a faster convergence rate
of the reconstruction algorithm. Nevertheless, finding the infimum and supremum over the whole frequency space $(\kvec ,\omega)$ implies difficult calculations. This is why a constraint or periodicity
is introduced for the functions in \eqref{Lamba} in order to find the values of the extrema $\Lambda_-$ and $\Lambda_+$. This periodicity can be found by defining a volumetric period
and we refer to \cite[Eq.(4.60) and Table 4.4] {Mujica99}   for the definition of this volumetric period and the values of the frame bounds
$A$ and $B$ for different values of the discretization parameters
$a_0$, $c_0$ and $\theta_0$. The resolution grid of the speed $c_0$ and orientation $\theta_0$ parameters must also be tuned to the GCM wavelet, but this is planned for a future work.

%Figure 4

\begin{figure}[t]
\begin{center}
\includegraphics[width=6cm,height=5.8cm]{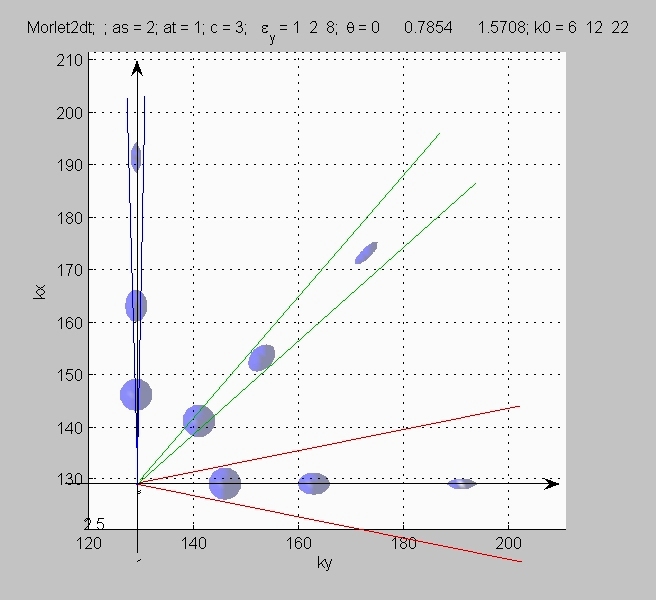}
\hspace{0.6cm}
\includegraphics[width=6cm,height=5.8cm]{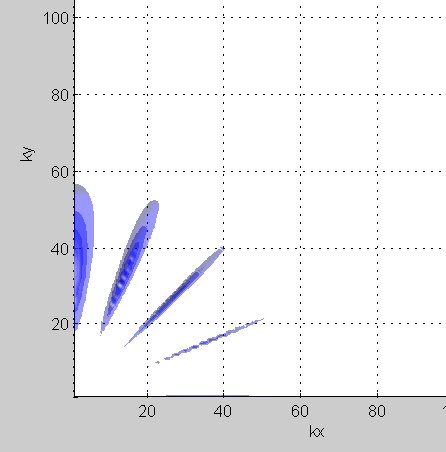}
\caption[]{\label{Morlet-vs-GCM2DT-2D}  \small{
Comparison of aperture selectivity tuning between Morlet and GCM. The figure shows that both wavelets are easily rotated. But, on panel (a), the Morlet wavelet behaviour, w.r.t. a tuning of its central wave-vector $k_0$, is awkward~: a change of $k_0$ induces a change in its opening angle $\alpha$, but the wavelet moves on its radius with $k_0$, then its radial extension considers a different frequency bandwidth each time. On the contrary, panel (b) shows that on GCM, a variation of the aperture does not imply any displacement on its radius and that the radial extension is constant and easily adjustable.  {This is a major strength of GCM~: its radial stability, with a variation in aperture, and its ability to be radially adjusted, which are very important for object recognition and tracking by spectral signature}. For sake of clarity we show the result for several rotations of GCM.}
}
\end{center}
\end{figure}

\subsection{Analysis algorithm}

For the analysis, we have to discretize the continuous CWT given in \eqref{eq:2DTCWT}. This means, given a sequence of $N$ frames,
corresponding the time variables $\tau_{i}, i=1,\ldots,N$,
 we take the product of the whole sequence FFT with each speed-tuned wavelet FFT, then we apply an inverse FFT. 
This yields  the wavelet transform  of the  whole sequence, ${\mathcal W}_\psi(\vec{b},\tau_{i}, \theta ; a_s,a_t,c_{j})$,  
with all the parameters suitably discretized, in particular $c=c_{j}$ and $\tau=\tau_{i}$.
We then compute the energy density of the $i^{\rm th}$ frame $|{\mathcal W}_\psi(c_{j},\tau_{i})|^2$,  taken as a function of  speed $c_{j}$ only, all the other parameters being fixed.
Then, selecting a frame or   a group of $N_0$ frames among the $N$ frames of the sequence,
we compute the total energy, $E_{\rm tot}(c_{j})$, of the pixels of the selected frames $N_0$:
\begin{equation}
E_{\rm tot}(c_{j})= \sum_{i\in N_0} \sum_{\bvec_{nm}}   |{\mathcal W}_\psi(\bvec_{nm},c_{j},\tau_{i})|^2.
\end{equation}
In this relation, $\bvec_{nm}$ denotes a discretized version of the $\bvec$-plane, the summation running over all pixels of the frame.
Note this expression is analogous to   the speed-orientation density described in \cite[Eq.(40)]{Mujica2000} or \cite[Eq.(10.60)]{Antoine04b},
except that here we sum over \emph{all} pixels of the frame (i.e., $\mathcal{B}= \Rbb^2$), because we want a \emph{global} energy density.

Then we study the curve $f(c_{j}):=E_{\rm  tot}(c_{j})$. This curve will go through a maximum, $v_m$, when the speed of the tuned wavelet $c_{j}$ matches the real speed  $v_r$ of
the object (a travelling 2D Gaussian here), see an example in Fig. \ref{Capture_spectra_and_Energy_curve_for_GCM}(b).
Several algorithms could be used to optimize the search for the maximum of the energy. This could be done, for example, by using a dichotomy rather than sweeping the whole range of velocity tunings,
or by using the Nelder-Mead algorithm as in \cite{Mujica2000}.
%
%
%
%=============================
\section{Experimentation}
%=============================
{We first   recall the results obtained in speed computation \cite{Brault05} between the Morlet speed-tuned CWT and the Optical Flow:}

Total computation speed (Xeon bi-processor at 2.4 gHz):

(1) MTSTWT with 3 wavelets tuned to 3 speeds (3,6,10 pixels/fr) on a $360\times 240\times 8$ sequence block (Tennis sequence) at the highest resolution of scale:
 $t_{\rm  MTSTWT}({1200{\rm  ms}})+ 3 \times {IFFT3D}(3 \times 380$  ms)  = 2.4 s.

(2) Fast Optical Flow with wavelets \cite{Bernard99.0} between two frames and 4 different resolutions~: $t_{\rm OF}$ =10 s.

In Fig. \ref{Morlet-vs-GCM2DT-2D}, we show the orientations that the Morlet and the GCM wavelets can take for $0\leqslant \theta \leqslant +\pi/2$. This figure speaks by itself.
 Morlet is shown with three orientations. In fact this figure illustrates the major problem of aperture adjustment with the Morlet wavelet : its opening angle is mainly trimmed
 through a variation of its central wave-vector $k_0$. And the larger the value of $k_0$, the smaller the aperture of the directional Morlet. But the major drawback of this adjustment is that the
frequency bandwidth that Morlet can capture also increases with the value of $k_0$. And because this bandwidth corresponds to the \textit{spectral signature} of the moving object, it can become
 difficult to adjust the aperture of the filter during object tracking. The effect of the anisotropy parameter $\epsilon$ does not have this drawback. But its effect is relatively limited to a variation of
the aspect-ratio of the wavelet along a specific direction. Here, we have also modified (increased) the value of $\epsilon$ in the $k_y$ direction (thus resulting in a elongation in the $k_x$ direction).
The displacement with $k_0$ is large enough to see, on the same radius, all the shapes that Morlet takes if we increase the couple $(k_{0},\epsilon)$. We show 3 couples $k_{0}=\{6,12,22\}$ and
 $\epsilon=\{1,2,8\}$.
For small values of $k_{0}$ and/or $\epsilon$, the aperture is very large. This aperture decreases by increasing $k_{0}$ and $\epsilon$. But as $k_{0}$ increases, the Morlet wavelet
moves away  from the Fourier center (0,0), along its radius. This makes very difficult to adjust the spatial positioning of this wavelet w.r.t. a change in aperture selectivity. 
 
With GCM, on the contrary,  the couple orientation/aperture is extremely simple to adjust. We show 5 orientations $\theta= \{0 \; \textrm{to} \; \pi/2\}$ with GCM tuned to aperture
 $\alpha=\{\pi/256 \; \textrm{to} \; \pi/16\}$. This proves the very good aperture selectivity of GCM.  {But it also proves its very good stability in radial expansion and position 
with a variation of aperture. These last characteristics, together with the adjustability of GCM in radial expansion and position, prove the real superiority of GCM on the 2D+T Morlet 
for object recognition and tracking by spectral signature.

%Figure 5
%
\begin{figure}[t]
\begin{center}
\includegraphics[width=7.2cm,height=6cm]{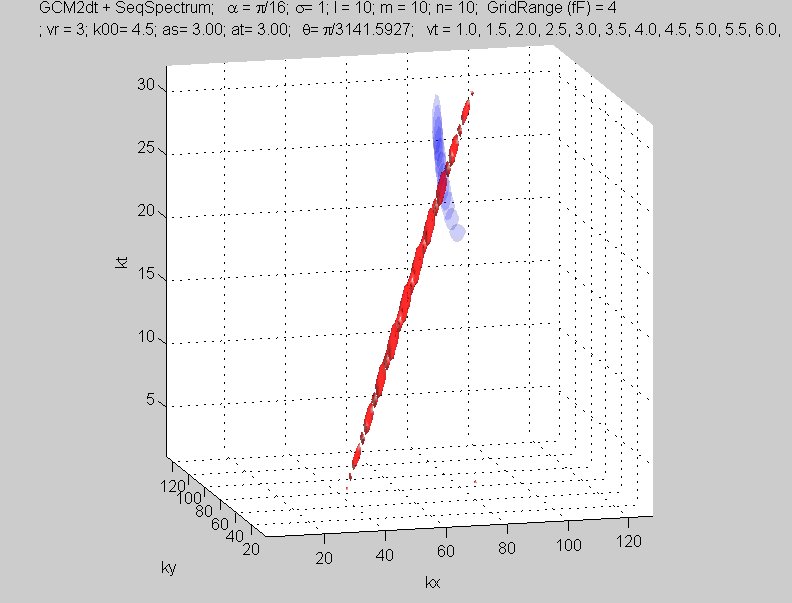}
\hspace{0.2cm}
\includegraphics[width=7.2cm,height=6cm]{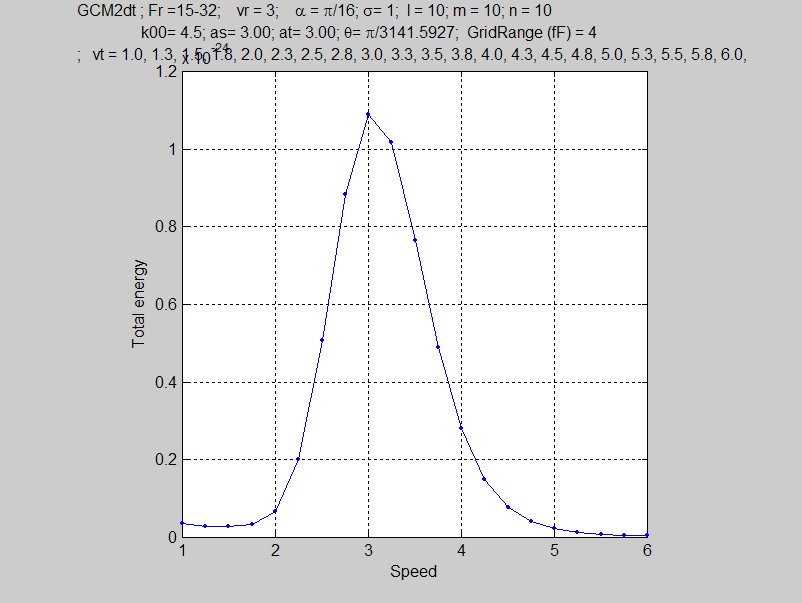}
\caption[]{\label{Capture_spectra_and_Energy_curve_for_GCM} \small
Demonstration of speed capture in the Fourier domain with $v_r = 3$.
(a)  {The spectrum of the sequence ``travelling Gaussian'' (in red) intersects the speed-tuned GCM wavelets hyperbola-like family (in blue)}.
(b) The energy sum plotted w.r.t wavelet speed tuning ($c$). The maximum is exactly reached for $c =v_{r}=$ 3 pixels/fr.}
\end{center}
\end{figure}
%
%Figure 6
%
\begin{figure}[h!]
\begin{center}
\includegraphics[width=7.2cm,height=6cm]{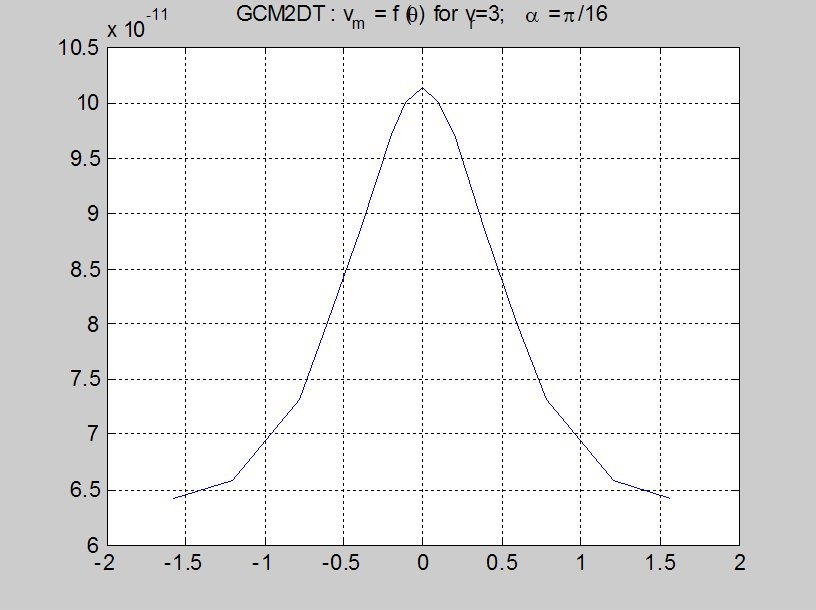}
\caption[]{\label{Speed_capture_GCM2_curve} \small Result~: Performances of GCM in directional speed capture:
we plot the curve $v_{m} $ vs. $\theta_{wav}$ for $-\pi/2 \leqslant \theta \leqslant +\pi/2$ for an aperture $\alpha=\pi/16$ and $v_r=3$.
The correct speed is captured when the wavelet orientation ($\theta=0$) exactly corresponds to the spectrum orientation (OX).
}
\end{center}
\end{figure}

%
%================================

We now perform a comparison between the 2D+T Morlet and the 2D+T GCM wavelets. We use a test sequence of $128 \times 128 \times 32$, that includes the motion, at constant speed, 
of a 2D non-symmetrical Gaussian.
The angles of the Gaussian and of its trajectory are varied (along OX, at  45$^\circ$ and along OY).

The first experiments have been done with an isotropic Gaussian travelling at constant speed in the plane. We checked that speed capture is as good with GCM as with Morlet. 
But this is not the real interest of this study.
 We then modified the Gaussian to exhibit a strong anisotropy in the direction OY in the spatial plane.
For this we took a large $\sigma_y$ (i.e., in the OY direction) and a small $\sigma_x$. The larger $\sigma_y$, the narrower its spectrum along the $k_x$ direction in the Fourier domain.
Thus we are able to test the angular, or more exactly, the aperture selectivity of the wavelet, i.e. its accuracy in directional speed capture.
The orientation of the object will also be its spectral ``signature'' and will enable to ``capture'' it in a sequence and to assign it its speed (Fig. \ref{Capture_spectra_and_Energy_curve_for_GCM}).

In the 3D Fig.\ref{Capture_spectra_and_Energy_curve_for_GCM}(a), we can observe the spectrum of the travelling Gaussian, oriented along OX (OY in the direct space) and of large $\sigma_y$,
 and travelling (along OX in the direct space) at a speed $v_r=$ 3 pixels/fr.
The figure shows the ``interception'' of the spectrum by the hyperbola-like family of GCM wavelets tuned to speeds between 1 and 6 pixels/fr.
The aperture is $\alpha=\pi/16$ and the spatial and temporal scales are $a_s=3$ and $a_t=3$. By looking again at Fig. \ref{Morlet-vs-GCM2DT-2D}, panel (a), it is obvious that in the same conditions,
a change of aperture of the Morlet wavelet would result in a displacement on its radius, thus in a change of spatial position and scale. The poor angular selectivity of Morlet, together with speed-tuning,
thus makes the directional capture of the spectrum very difficult. This does not happen with GCM. This makes it very efficient in capturing the signal speed at any angle within a constant 
 narrow conical aperture
$\alpha=\pi/16$ and not outside it. We will also show further that GCM can achieve much narrower apertures in speed-capture.
%===============================
%

In Fig. \ref{Speed_capture_GCM2_curve}, we plot the curve $v_{m}$ vs. $\theta_{wav}$ for $-\pi/2 \leqslant \theta \leqslant +\pi/2$, with GCM.
The correct speed is captured when the wavelet orientation ($\theta=0$) exactly corresponds to the spectrum orientation (OX).
This proves the good angular selectivity of GCM, that could not be reached with Morlet, and its efficiency
to detect the correct speed of the sequence in a very narrow angular aperture and not elsewhere.%}}

%
%================================
%Figure 7

\begin{figure}[h!]
\begin{center}
\includegraphics[width=6cm,height=4.8cm]{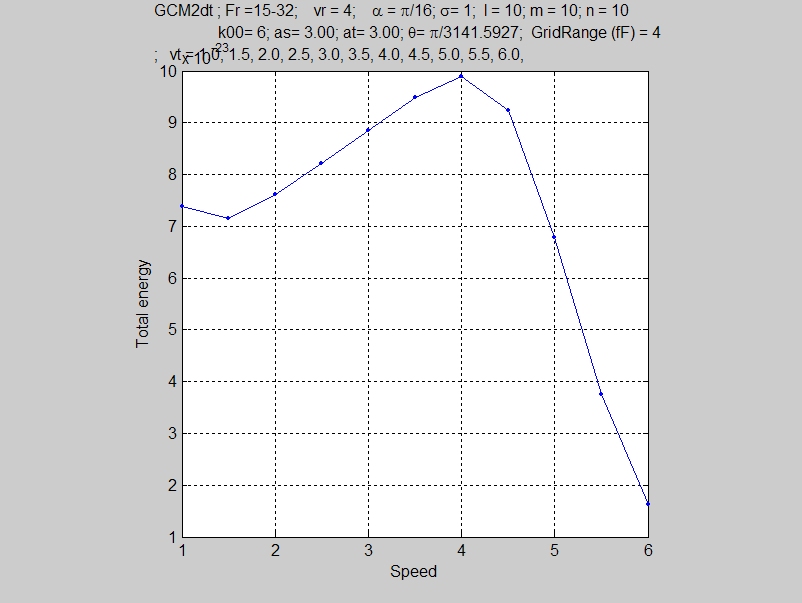}
\hspace{0.2cm}
\includegraphics[width=6cm,height=4.8cm]{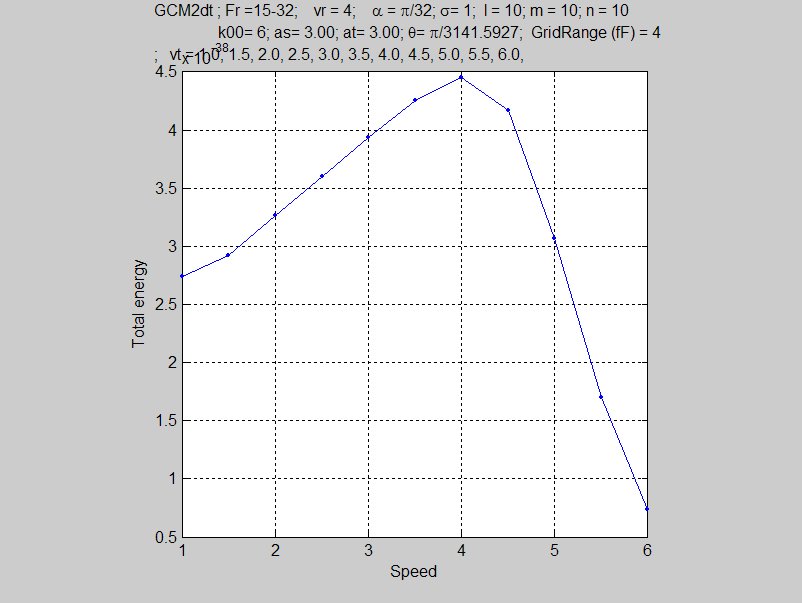}\\
\vspace{0.2cm}
\includegraphics[width=6cm,height=4.8cm]{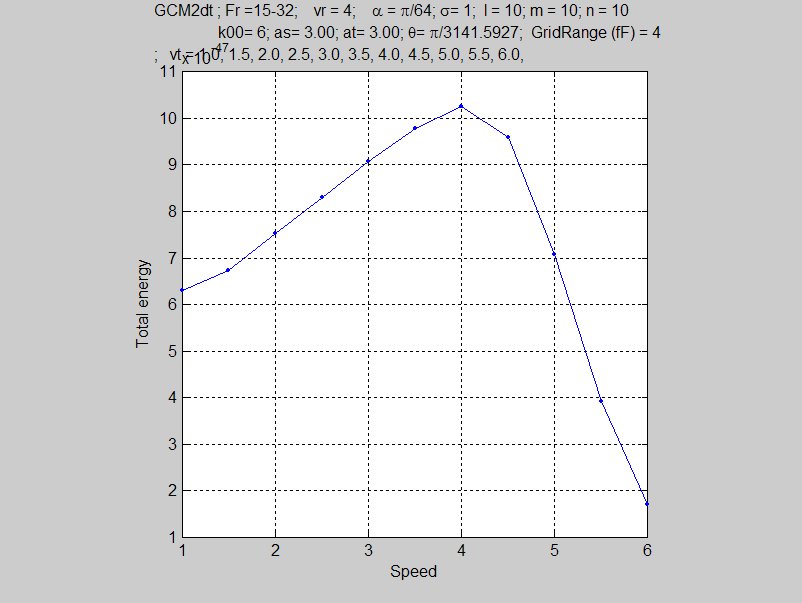}
\hspace{0.2cm}
\includegraphics[width=6cm,height=4.8cm]{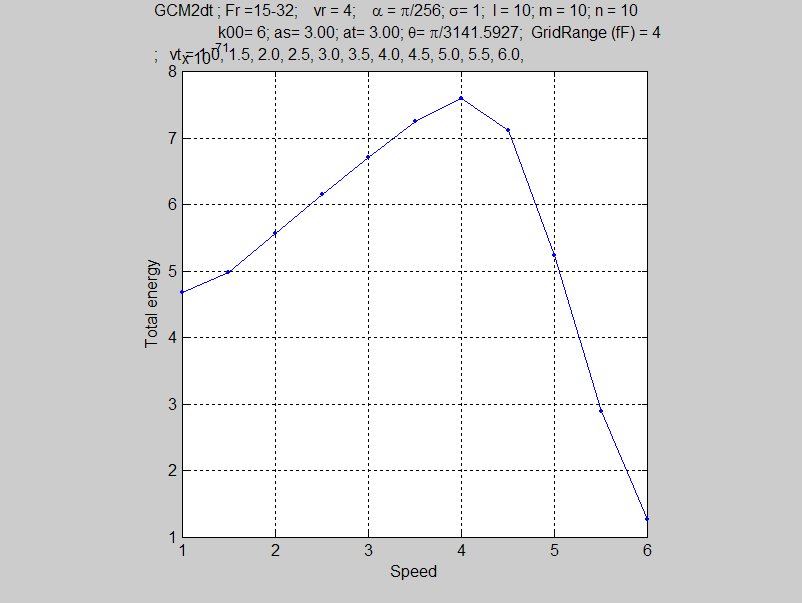}
\caption[]{\label{Test_GCM_4_apertures} {\small
  This figure shows a speed detection for an object speed $v_r=4$, like in Fig. \ref{Capture_spectra_and_Energy_curve_for_GCM}(b), but for a more and more narrow wavelet aperture
$\alpha$,} respectively (from $(a)$ to $(d)$)~: $\pi/8, \pi/16, \pi/64, \pi/256$. Here the GCM parameters are $\sigma=1, {l=m=10}, k_0=6$ and $a_s=a_t=3$.
We observe a very good capacity of capture for a very narrow aperture  ($\pi/256=0.70$  deg).}
\end{center}
\end{figure}
%===============================================================

The final Fig. \ref{Test_GCM_4_apertures}  demonstrates the excellent performance of GCM during a speed capture
at $v_r=4$ and for increasingly  narrow conical apertures.
We have taken apertures of $\pi/8, \pi/16, \pi/64, \pi/256$ (0.70 deg). The capture conditions are those of
Fig. \ref{Capture_spectra_and_Energy_curve_for_GCM}, where the spectrum, oriented in the OX direction,
is narrow, and for a real speed $v_r=4$. GCM captures the spectrum for extremely weak apertures without moving in the Fourier space. This makes it very robust to scale initialization and object size detection and tracking,
which is one a the major drawbacks of speed capture with Morlet.
%
%
%--------------------------
\section{Conclusion}
%--------------------------
{In this paper, we have introduced a new tool, based on the redundant CWT,  for directional speed analysis in video sequences,
namely, a new speed-tuned wavelet, the Gaussian-Conical-Morlet (GCM). It is based on a 2D Gaussian-Conical genuine}
directional wavelet, on one hand, together with a temporal 1D Morlet wavelet, on the other hand, and it provides a very high directionally selective speed-tuned wavelet.
This substantially improves the characteristics of the 2D+T Morlet wavelet previously used in spatio-temporal speed-based approaches of motion estimation and tracking. It thus reinforces the efficiency of
the CWT tool for speed and motion tracking, and in particular its inherent robustness to noise, occlusions and local illumination variations as well as its efficiency to long dependence analysis.
{In this work we have proved the extreme efficiency of this tool in directional speed selectivity (the aperture of the conical wavelet) down to angle apertures of less than 1 degree ($\pi/256$) but also its capacities of radial stability and adjustment, w.r.t. aperture variation, that make it a much more powerful tool than the 2D+T Morlet wavelet for spectral signature recognition and tracking.}
An initialization of the speed capture has also been proposed by turning the GCM wavelet to a simple low-pass directional filter centered in the middle of the Fourier plane.
These two contributions will be pursued with,  among other topics, applications in the domain of object and motion tracking in video sequences and in the domain of feature retrieval.
%

%==================================

\end{document}